\RequirePackage{amsmath} 
\documentclass[runningheads]{llncs} 
\usepackage[T1]{fontenc}
\usepackage{graphicx}
\usepackage{hyperref}
\hypersetup{
    colorlinks,
    linkcolor={red!50!black},
    citecolor={blue!50!black},
    urlcolor={blue!80!black}
}
\usepackage{orcidlink}
\usepackage{color}

\usepackage{amsmath,amsfonts,amssymb}
\usepackage{xfrac}
\usepackage{mathtools}
\usepackage{physics}
\usepackage[normalem]{ulem}

\usepackage{nicematrix}
\usepackage{quantikz}
\usepackage[ruled,vlined,linesnumbered,nofillcomment]{algorithm2e}
\usepackage[caption=false]{subfig}
\usepackage{makecell}
\usepackage{multirow}
\usepackage{wrapfig}
\usepackage{listings} 
\usepackage{float}
\usepackage{tabularray}

\usepackage{enumitem}
\usepackage[capitalise]{cleveref}
\usepackage{refcount}
\usepackage[textsize=scriptsize]{todonotes}

\let\oldparagraph=\paragraph
\renewcommand\paragraph[1]{\oldparagraph{#1.}}

\newcommand{\arxiv}[2]{#1} 

\newcommand{\transpose}{\intercal}
\newcommand{\mymatrix}{pmatrix*}
\DeclareMathSymbol{\smin}{\mathbin}{AMSa}{"39}
\newcommand{\CX}{\mathit{CX}}
\newcommand{\normsymb}{\nu}

\crefname{line}{line}{lines} 

\newcommand{\fname}{\textsc} 
\SetFuncArgSty{textnormal}
\SetArgSty{textnormal}
\SetKwProg{Fn}{def}{}{}
\SetFuncSty{textsc}
\SetVlineSkip{0pt}
\DontPrintSemicolon

\SetKwComment{Comment}{$\triangleright$ }{}
\SetCommentSty{mycommfont}
\SetKw{Or}{\textnormal{\textbf{or}}}
\SetKw{And}{\textnormal{\textbf{and}}}
\newcommand{\follow}[2]{#2[#1]}
\newcommand{\makenode}[1]{\fname{MakeNode}(#1)}
\newcommand{\ddvar}[1]{\textsf{var}(#1)}
\newcommand{\edgeval}[1]{\textsf{val}(#1)}

\makeatletter
\patchcmd{\@algocf@start}
  {-1.5em}
  {0pt}
  {}{}
\makeatother

\definecolor{codegreen}{rgb}{0,0.6,0}
\definecolor{codegray}{rgb}{0.5,0.5,0.5}
\definecolor{codepurple}{rgb}{0.58,0,0.82}
\definecolor{backcolour}{rgb}{0.95,0.95,0.92}
\lstdefinestyle{myCstyle}{
    language=C,
    backgroundcolor=\color{backcolour},   
    commentstyle=\color{codegreen},
    keywordstyle=\color{magenta},
    numberstyle=\tiny\color{codegray},
    stringstyle=\color{codepurple},
    basicstyle=\ttfamily\footnotesize,
    breakatwhitespace=false,         
    breaklines=true,                 
    captionpos=b,                    
    keepspaces=true,                 
    numbers=left,                    
    numbersep=5pt,                  
    showspaces=false,                
    showstringspaces=false,
    showtabs=false,                  
    tabsize=2,
    morekeywords={bool,EVBDD,EVDD,QMDD}
}

\usetikzlibrary{patterns,shapes,arrows,fit,calc,positioning,automata}
\tikzset{unused/.style={pattern = north east lines}}
\tikzset{main/.style={draw,circle}} 
\tikzset{leaf/.style={draw,minimum width=1.2em,minimum height=1.2em}} 
\tikzset{e0/.style={draw,->,dotted,>=latex}}
\tikzset{e1/.style={draw,->,>=latex}}
\tikzset{XOR/.style={draw,circle,append after command={
        [shorten >=\pgflinewidth, shorten <=\pgflinewidth,]
        (\tikzlastnode.north) edge (\tikzlastnode.south)
        (\tikzlastnode.east) edge (\tikzlastnode.west)
        }
    }
}

\newcommand{\leafsymb}{1}

\renewcommand{\leafsymb}{\top}

\begin{document}
\title{Q-Sylvan: A Parallel Decision Diagram Package for Quantum Computing}
\titlerunning{ }
%
\author{Sebastiaan Brand\orcidlink{0000-0002-7666-2794} \and
Alfons Laarman\orcidlink{0000-0002-2433-4174}}
\authorrunning{S. Brand and A. Laarman}
%
\institute{Leiden Institute of Advanced Computer Science, Leiden University, The Netherlands
\email{\{s.o.brand,a.w.laarman\}@liacs.leidenuniv.nl}}
\maketitle              
\begin{abstract}
As physical realizations of quantum computers move closer towards practical applications, the need for tools to analyze and verify quantum algorithms grows. Among the algorithms and data structures used to tackle such problems, decision diagrams (DDs) have shown much success. However, an obstacle with DDs is their efficient parallelization, and while parallel speedups have been obtained for DDs used in classical applications, attempts to parallelize operations for quantum-specific DDs have yielded only limited success. In this work, we present an efficient implementation of parallel edge-valued DDs, which makes use of fine-grained task parallelism and lock-free hash tables. Additionally, we use these DDs to implement two use cases: simulation and equivalence checking of quantum circuits. In our empirical evaluation we find that our tool, \mbox{Q-Sylvan}, shows a single-core performance that is competitive with the state-of-the-art quantum DD tool MQT DDSIM on large instances, and moreover achieves parallel speedups of up to $\times 18$ on 64 cores.

\end{abstract}

\section{Introduction}
\label{sec:intro}
Quantum computing is an emerging technology that aims to provide computational speedups on problems in areas such as cryptography~\cite{shor1994algorithms}, finance~\cite{orus2019quantum}, and optimization~\cite{harwood2021formulating}, as well as on problems in quantum physics~\cite{huang2022provably} and quantum chemistry~\cite{cao2019quantum}.
As the number of available qubits on quantum chips grows, and the field moves closer towards practical applications, so grows the need for tools to analyze and verify quantum circuits.

In this paper, we focus on two tasks specifically: simulation and equivalence checking of quantum circuits.
Although these are computationally hard problems~\cite{aaronson2010bqp,janzing2005non,ji2009non,tanaka2010exact},
much like problems in classical verification, heuristic algorithms and data structures can significantly help with the scalability of these tasks.
One particular data structure that has seen a lot of success in the classical evaluation of quantum circuits is decision diagrams (DDs).
While different types of decision diagrams have been proposed in the context of quantum computing~\cite{viamontes2004high,samoladas2008improved,miller2006qmdd,zulehner2018advanced,tsai2021bitslicing,wei2022accurate,vinkhuijzen2023limdd,sistla2023symbolic}, edge-valued decision diagrams (EVDDs)~\cite{tafertshofer1997factored} with complex edge values~\cite{miller2006qmdd,zulehner2018advanced} have been shown to be very useful in practice.

However, a particular obstacle with decision diagrams is that computations on them are hard to efficiently parallelize~\cite{vandijk2013multicore}. And although the decision diagram library Sylvan~\cite{vandijk2017sylvan} has shown good speedups for DDs without edge values, the storage and handling of floating-point values in EVDDs adds additional complications. This has thus far yielded limited scalability of EVDDs for quantum computing applications, with speedups of up to $\times 3$ using 32 cores on the simulation of random circuits~\cite{hillmich2020concurrency} and $\times 2\textrm{--}3$ using 16 cores on Grover circuits~\cite{li2024parallelizing}.

In this paper, we present an efficient implementation of parallel EVDDs, building on the parallel decision diagram library Sylvan~\cite{vandijk2017sylvan}, which before now only supported decision diagrams without edge values, and we address several obstacles regarding floating-point values in DDs. We also provide implementations for two use cases: quantum circuit simulation and quantum circuit equivalence checking, both supporting the full set of standard quantum gates of Open QASM~2.0~\cite{cross2017open}.
We evaluate the performance of our resulting tool, \mbox{Q-Sylvan},\footnote{Available online at \url{https://github.com/System-Verification-Lab/q-sylvan} under the Apache-2.0 license.} against several other recent tools on a large set of benchmarks. We show that Q-Sylvan is capable of obtaining speedups of up to $\times 7$ on 8 cores and up to $\times 18$ on 64 cores, while also providing a single-core performance that is competitive with the state-of-the-art DD-based quantum circuit simulator MQT DDSIM~\cite{zulehner2018advanced} on large instances.
The contributions of this paper are summarized as follows:
\begin{enumerate}
    \item An efficient implementation of parallel EVDDs.
    \item The implementation of two use cases using these EVDDs: simulation and equivalence checking of quantum circuits.
    \item An evaluation against state-of-the-art tools on a large benchmark set.
\end{enumerate}

\section{Quantum computing and EVDDs}
For the convenience of the reader, we briefly explain the necessary basics of quantum computing and the role of decision diagrams in this context.

While there are many interesting intricacies in the mathematics of quantum computing, for the purposes of this paper, a high-level description of \emph{quantum states} and \emph{quantum gates} will be sufficient.
The state of $n$ quantum bits (qubits) can be described by a vector in $\mathbb{C}^{2^n}$, and can also be seen as a function \mbox{$\psi : \{0,1\}^n \to \mathbb{C}$}. Similarly, a quantum gate (i.e. a linear transformation that maps states to states) that acts on $n$ qubits can be described by a matrix in $\mathbb{C}^{2^n \times 2^n}$ (or a function $U : \{0,1\}^{2n} \to \mathbb{C}$). The effect of a gate on a state can be computed through matrix-vector multiplication. 
A \emph{quantum circuit} is a sequence of quantum gates, and can be classically simulated through repeated matrix-vector multiplication. 

To store these exponentially large vectors and matrices compactly, people have turned to decision diagrams, a data structure for which it is well known how to compute matrix multiplications~\cite{fujita1997multi}. In this work we opt to use edge-valued DDs~\cite{miller2006qmdd}, which are also called quantum multi-valued DDs (QMDDs)~\cite{zulehner2018advanced}.

\begin{wrapfigure}[8]{r}{1.9cm}
    \centering
    \scalebox{.9}{\begin{tikzpicture}[auto, thick,node distance=1.cm,inner sep=1.5pt]
    \newcommand\dist{.33cm}
    \newcommand\sep{.7cm}
    \newcommand\sepp{0.15cm}
    \newcommand\seppp{-.15cm}
    \node[] (0) [] {};
    \node[main] (x0) [node distance=.65cm,below of=0] {$x_0$};
    \node[main] (x1)  [below right = .9*\dist and 1.2*\sepp of x0] {$x_1$};
    \node[main] (x21) [below left  = 3.4*\dist and 1.2*\sepp of x0] {$x_2$};
    \node[main] (x22) [below = \dist of x1] {$x_2$};
    \node[leaf] (l)   [below = 5.4*\dist of x0] {$\leafsymb$};
    \draw[e1] (0) to (x0);
    \draw[e0,bend right=20] (x0) to (x21);
    \draw[e1] (x0) to (x1);
    \draw[e0,bend right=20] (x1) to (x22);
    \draw[e1,bend left=20] (x1) to node [] {\hspace{.5mm}$3$} (x22);
    \draw[e0,bend right=20] (x21) to (l);
    \draw[e1,bend left=10] (x21) to node [right,pos=0] {$\smin2$} (l);
    \draw[e0,bend right=10] (x22) to (l);
    \draw[e1,bend left=20] (x22) to node [right,pos=.5] {\hspace{.5mm}$i$} (l);
\end{tikzpicture}}%
\end{wrapfigure}
An EVDD is a rooted, directed, acyclic graph whose edges have values associated with them. For our purposes these values are complex numbers. 
Each node $v$ in an EVDD has a variable $\ddvar{v} = x_i$, and two outgoing edges. 
For an EVDD that encodes an $n$-qubit state, we use the variables $\{x_0, \dots, x_{n-1}\}$. 
EVDDs are always ordered, i.e. on every path the variables are encountered in the same order $x_0 \prec x_1 \prec \dots \prec x_{n-1}$, although variables may be skipped.
Every path through an EVDD corresponds to a single entry in the vector it encodes, with that value being equal to the product of the edge values on that path.
As an example, the EVDD on the right encodes the vector
$\vec \psi = \begin{\mymatrix} 1 & \smin 2 & ~1 & \smin 2 & ~1 & ~i & ~3 & ~3i \end{\mymatrix}^\transpose$. The value $\psi(110)$, for example, can be read from the EVDD by following the 1 (solid) edges for $x_0$ and $x_1$, and the 0 (dashed) edge for $x_2$, obtaining $1 \cdot 3 \cdot 1 = 3$.
Algorithms on DDs are typically defined recursively. Two examples are given in \Cref{alg:evdd-plus,alg:evdd-mult}.

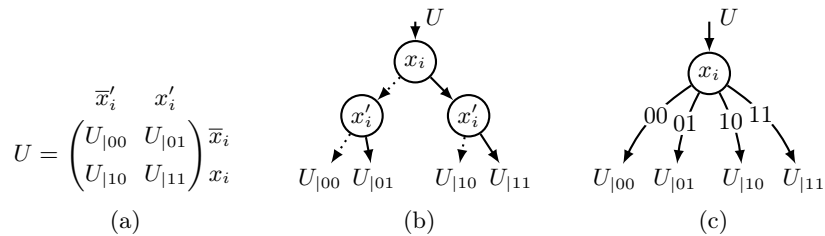
\begin{figure}[t]
    \centering
    \captionsetup{captionskip=-3mm}
    \subfloat[]{%
        \begin{minipage}[b][2.0cm][t]{3.8cm}
            \centering
            \renewcommand{\arraystretch}{1.3}
            \setlength\arraycolsep{3pt}
            $U = 
            \begin{pNiceMatrix}[first-row,last-col]
            \overline{x}_i' & x_i' & \\
            U_{|00} & U_{|01} & \overline{x}_i\\
            U_{|10} & U_{|11} & x_i
            \end{pNiceMatrix}
            $
            \end{minipage}
        \label{fig:matrix-example}%
    }%
    \subfloat[]{%
        \begin{minipage}[b][3.0cm][t]{3.8cm}\centering\begin{tikzpicture}[auto, thick,node distance=1.cm,inner sep=1.5pt]
    \tikzstyle{temp}=[main, inner sep=0, minimum size=.55cm] 
	\node[] (0) [label=right:{$U$}] {};
	\node[temp] (1) [node distance=.6cm,below of=0] {$x_i$};
	\node[temp] (2) [below left=.31cm and 0.3cm of 1] {$x_i'$};
	\node[temp] (3) [below right=.31cm and 0.3cm of 1] {$x_i'$};
	\node[] (00) [below left=.43cm and 0cm of 2] {$U_{|00}$};
	\node[] (01) [below right=.43cm and -0.4cm of 2] {$U_{|01}$};
	\node[] (10) [below left=.43cm and -0.4cm of 3] {$U_{|10}$};
	\node[] (11) [below right=.43cm and 0cm of 3] {$U_{|11}$};
	\draw[e1] (0) to (1);
	\draw[e0] (1) to (2);
	\draw[e1] (1) to (3);
	\draw[e0] (2) to (00);
	\draw[e1] (2) to (01);
	\draw[e0] (3) to (10);
	\draw[e1] (3) to (11);
\end{tikzpicture}\end{minipage}
        \label{fig:matrix-evbdd}%
    }
    \subfloat[]{%
        \begin{minipage}[b][3.0cm][t]{3.8cm}\centering\begin{tikzpicture}[auto, thick,inner sep=1.5pt,edgelabel/.style={fill=white,rounded corners=4pt,inner sep=1.5pt}]
    \tikzstyle{temp}=[main, inner sep=0, minimum size=.55cm] 
	\node[] (0) [label=right:{$U$}] {};
	\node[temp] (1) [below=4mm of 0] {$x_i$};
	\node[] (00) [below  left=10mm and 7mm of 1] {$U_{|00}$};
	\node[] (01) [below  left=10mm and -1mm of 1] {$U_{|01}$};
	\node[] (10) [below right=10mm and -1mm of 1] {$U_{|10}$};
	\node[] (11) [below right=10mm and 7mm of 1] {$U_{|11}$};
	\draw[e1] (0) to (1);
	\draw[e1,bend right=15] (1) to node[edgelabel,pos=0.25,xshift=-4mm] {00} (00);
	\draw[e1,bend right=10] (1) to node[edgelabel,pos=0.25,xshift=-3mm] {01} (01);
	\draw[e1,bend  left=10] (1) to node[edgelabel,pos=0.6,xshift=-3mm] {10} (10);
	\draw[e1,bend  left=15] (1) to node[edgelabel,pos=0.6,xshift=-4mm] {11} (11);
\end{tikzpicture}\end{minipage}
        \label{fig:matrix-qmdd}%
    }%
    \vspace{-2mm}
    \caption{A $2^n \times 2^n$ matrix \protect\subref{fig:matrix-example} can be recursively encoded in a DD with two children per node \protect\subref{fig:matrix-evbdd}, as in EVDD~\cite{tafertshofer1997factored}, or four children per node \protect\subref{fig:matrix-qmdd}, as in QMDD~\cite{miller2006qmdd}.
    }
    \label{fig:qmdd-evbdd-matrix}
\end{figure}

\Cref{fig:qmdd-evbdd-matrix} shows how DDs represent matrices. EVDDs do so by reinterpreting the matrix as a function $f(\vec x, \vec x')$, where $\vec x$ indexes rows and $\vec x'$ columns (\cref{fig:matrix-example}). The variables in $\vec x$ and $\vec x'$ are then interleaved to enable recursive descent on matrix quadrants (\cref{fig:matrix-evbdd}). In contrast, a QMDD~\cite{miller2006qmdd} is refined for matrices, as \Cref{fig:matrix-qmdd} shows.
Because a QMDD can be translated into an EVDD in linear time and vice versa, they are effectively the same data structure~\cite{fargier2014knowledge}.

Other types of DDs have also been used within quantum computing, such as multi-terminal binary DDs (MTBDDs)~\cite{viamontes2004high,samoladas2008improved}, context-free-language ordered BDDs (CFLOBDDs)~\cite{sistla2023symbolic}, local invertible map DDs (LIMDDs)~\cite{vinkhuijzen2023limdd}, and local invertible map tensor DDs (LimTDDs)~\cite{hong2025limtdd}. EVDDs appear to strike a good balance between compression and practical efficiency: compared to MTBDDs, they benefit from better compression~\cite{zulehner2018advanced}, while compared to LIMDDs and CFLOBDDs, they are practically more efficient on a wider array of circuits~\cite{sistla2023symbolic,vinkhuijzen2023efficient}.

\section{EVDD implementation}
\label{sec:evdd-implementation}

Since our goal is to provide an efficient implementation of parallel EVDDs, we choose to implement these on top of the parallel decision diagram library Sylvan~\cite{vandijk2017sylvan}. Sylvan contains implementations of several different types of decision diagrams, among which are multi-terminal binary decision diagrams (MTBDDs), list decision diagrams (LDDs), and zero-suppressed decision diagrams (ZDDs). However, Sylvan does not yet support any decision diagrams with edge values.

\begin{algorithm}[t]
\SetKwFunction{add}{Plus}
\Fn{\add{EVDD $A$, EVDD $B$}}{
    \vspace{-1.25em}\Comment*[r]{assuming $\ddvar{A} = \ddvar{B}$}

    \BlankLine
    \If{$A$ and $B$ are terminals}{
        \Return $\edgeval{A} + \edgeval{B}$
    }

    \BlankLine
    \lIf{$R \gets$ \textnormal{cache[\add,$A,B$]}}{
        \Return{$R$}
    }
    \vspace{-1.25em}\Comment*[r]{Memoization to avoid}
    \Comment*[r]{recomputing~~~~~~~~~}
    \vspace{-1.25em}

    \BlankLine

    $R_0 \gets$ \add{$\edgeval{A} \cdot \follow{0}{A}$, $\edgeval{B} \cdot \follow{0}{B}$} \; \label{line:rec-add0}
    $R_1 \gets$ \add{$\edgeval{A} \cdot \follow{1}{A}$, $\edgeval{B} \cdot \follow{1}{B}$} \; \label{line:rec-add1}
    $R \gets \makenode{\ddvar{A}, R_0, R_1}$ \;

    \BlankLine
    \textnormal{cache[\add,$A,B$]} $\gets R$\; 
    \BlankLine
    
    \Return{$R$}
    \scalebox{.90}{\begin{tikzpicture}[node distance=10mm, inner sep=1.5pt, auto, thick, overlay, xshift=5.1cm, yshift=1.5cm] 
	\node[] (0a) [] {};
	\node[main] (1a) [node distance=7mm,below of=0a] {$x_i$};
	\node[] (2a) [below left  = 4mm and -2mm of 1a] {$\follow{0}{A}$};
	\node[] (3a) [below right = 4mm and -2mm of 1a] {$\follow{1}{A}$};
	\draw[->] (0a) to node [left,pos=.3] {$A$} (1a) ;
	\draw[->,dashed] (1a) to (2a);
	\draw[->] (1a) to (3a);

	\node[] (plus) [right = .3cm of 1a] {$+$};

	\node[main] (1b) [right = .3cm of plus] {$x_i$};
	\node[] (0b) [node distance=7mm, above of=1b] {};
	\node[] (2b) [below left  = 4mm and -2mm of 1b] {$\follow{0}{B}$};
	\node[] (3b) [below right = 4mm and -2mm of 1b] {$\follow{1}{B}$};
	\draw[->] (0b) to node [left,pos=.3] {$B$} (1b) ;
	\draw[->,dashed] (1b) to (2b);
	\draw[->] (1b) to (3b);

	\node[] (equals) [right = .6cm of 1b] {$=$};

	\node[main] (1c) [right = 1.0cm of equals] {$x_i$};
	\node[] (0c) [node distance=7mm, above of=1c] {};
	\node[] (2c) [below left  = 4mm and -1.7mm of 1c] {$\follow{0}{A}$+$\follow{0}{B}$};
	\node[] (3c) [below right = 4mm and -1.7mm of 1c] {$\follow{1}{A}$+$\follow{1}{B}$};
	\draw[->] (0c) to node [left,pos=.3] {$R$} (1c) ;
	\draw[->,dashed] (1c) to (2c);
	\draw[->] (1c) to (3c);
\end{tikzpicture}}
}

\newcounter{linerecadd}
\setcounterref{linerecadd}{line:rec-add0}

\caption{Vector addition with EVDDs. Here $\edgeval{A}$ denotes the edge value of the root edge of $A$, and $\follow{0}{A}$ ($\follow{1}{A}$) denotes the 0 (1) child of $A$.}
\label{alg:evdd-plus}
\end{algorithm}
\begin{algorithm}[t]
\SetKwFunction{mult}{Multiply}
\Fn{\mult{EVDD $M$, EVDD $V$}}{
    \vspace{-1.25em}\Comment*[r]{assuming $\ddvar{M} = \ddvar{V}$}
    \BlankLine
    \If{$M$ and $V$ are terminals}{
        \Return $\edgeval{M} \cdot \edgeval{V}$
    }

    \BlankLine
    \lIf{$R \gets$ \textnormal{cache[\mult,$M,V$]}}{
        \Return $R$
    }

    \BlankLine
    $R_{00} \gets $\mult{$\edgeval{M[0]} \cdot \follow{00}{M},\follow{0}{V}$} \; \label{line:rec-mult00}
    $R_{01} \gets $\mult{$\edgeval{M[0]} \cdot \follow{01}{M},\follow{1}{V}$} \;
    $R_{10} \gets $\mult{$\edgeval{M[1]} \cdot \follow{10}{M},\follow{0}{V}$} \;
    $R_{11} \gets $\mult{$\edgeval{M[1]} \cdot \follow{11}{M},\follow{1}{V}$} \; \label{line:rec-mult11}
    \BlankLine
    $R_0 \gets \makenode{\ddvar{v},R_{00},R_{10}}$ \;
    $R_1 \gets \makenode{\ddvar{v},R_{01},R_{11}}$ \;
    \BlankLine
    $R \gets \fname{Plus}(R_0, R_1)$ \Comment*[r]{using \Cref{alg:evdd-plus}}

    \BlankLine
    \textnormal{cache[\mult,$M,V$]} $\gets R$\; 
    \BlankLine

    \Return $R \cdot \edgeval{M} \cdot \edgeval{V}$
}

\newcounter{linerecmult}
\setcounterref{linerecmult}{line:rec-mult00}
\caption{Matrix-vector multiplication with EVDDs.} 
\label{alg:evdd-mult}
\end{algorithm}

\paragraph{Floating-point equality}
A prominent hurdle in the implementation of DDs with real (or complex) edge values is handling floating-point values.
Floating-point computations infamously do not always yield exact solutions, e.g. $0.1 + 0.2$ might give $0.30000000000000004$, and so using exact floating-point equality would often prevent the merging of nodes, which in turn prevents the decision diagrams from staying compact. Therefore, we would like to consider edges with edge values that are very close to be equivalent. Specifically, we consider two floating-point values $a,b$ equivalent if $|a - b| < \delta$, with $\delta = 10^{\smin14}$, and we consider two complex values equivalent if the same inequality holds for both their real and imaginary components. Although this can introduce numerical errors, setting $\delta = 0$ has been shown empirically to allow for almost no node-merging~\cite{niemann2020overcoming}, and thus setting $\delta$ to a small but non-zero value appears to be a necessary evil of decision diagrams with floating-point edge values.
Alternative representations of real or complex values, such as algebraic representations, have their own shortcomings. For example, the algebraic representation proposed in~\cite{niemann2020overcoming} only works for a limited gate set (specifically $\{H,T,\CX \}$). And even though this gate set is technically universal for quantum computing, more general single-qubit rotation gates, which appear in ubiquitous quantum algorithms such as quantum approximate optimization algorithms (QAOA), variational quantum eigensolvers (VQE), and the quantum Fourier transform (QFT), can only be approximated by it. Specifically, an $m$-gate quantum circuit that contains general single-qubit rotation gates can be approximated up to an error of $\varepsilon$ by a circuit with $O(m\log(m/\varepsilon))$ gates~\cite{kitaev1997quantum,dawson2005solovay,selinger2012efficient}.



\paragraph{Storing edge values}
In order to efficiently recognize equivalent edge values, we store them in a hash table. To facilitate efficient concurrent access we use a hash table with atomic compare-and-swap operations based on~\cite{laarman2010boosting}. 
As mentioned above, due to the imperfection of floating-point arithmetic we cannot simply use floating-point equality to determine if two edge values are equivalent. Instead, when storing a complex edge value $c$ we hash a rounded version of $c$ to compute a bucket index, in which we store the (non-rounded) value of $c$. If the bucket is already occupied we compare the absolute values of the real and imaginary components to check if the stored value is equivalent to the value we want to store. A pseudocode description of this procedure is given in \Cref{alg:find-or-put}.

{
\begin{algorithm}[t]
\SetKwFunction{findorput}{FindOrPut}
\Fn{\findorput{$c$}} {
    \BlankLine
    $d.r \gets \fname{round}(c.r, \delta)$ \Comment*[r]{Round real and imaginary components}
    $d.i \gets \fname{round}(c.i, \delta)$ \;
    index $\gets \fname{hash}(d)$ \Comment*[r]{Compute hash based on rounded value}
    \While{\text{not found or put}}{
        \If{\textsc{CAS}(table[index], empty, $c$)} { \label{line:bucket-search}
            \vspace{-1.25em}
            \Comment*[r]{Store unrounded value with}
            \Return{index} \Comment*[r]{atomic compare-and-swap~~~}
        }\Else{
            $v \gets $ table[index] \;
            \If{ $|c.r - v.r| < \delta$ \And $|c.i - v.i| < \delta$}{
                \vspace{-1.25em}
                \Comment*[r]{If $\delta$-close, return}
                \Return{index} 
                \Comment*[r]{existing value~~~~}
            }\Else {
                index $\gets$ index + 1 \;
            }
        }
    }
}

\caption{Find-or-put a complex value $c$ in a hash table. $\delta$ is a configurable variable with a default value of $10^{-14}$.}
\label{alg:find-or-put}
\end{algorithm}
}

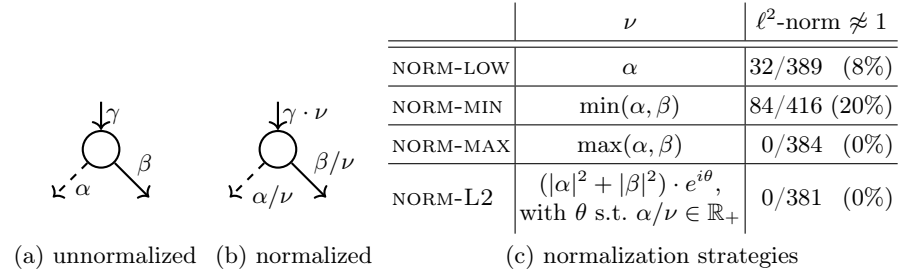
\begin{figure}[t]
    \centering
    \captionsetup{captionskip=0mm}
    \subfloat[unnormalized]{%
        \begin{minipage}[b][2.0cm][t]{2.4cm}\centering\begin{tikzpicture}[node distance=10mm, inner sep=1.5pt, auto, thick] 
    \node[] (0) [] {};
    \node[main] (1) [node distance=7mm,below of=0] {\phantom{$x_i$}};
    \node[] (2) [below left of=1] {};
    \node[] (3) [below right of=1] {};
    \draw[->] (0) to node [] {$\gamma$} (1) ;
    \draw[->,dashed] (1) to node [] {$\alpha$} (2);
    \draw[->] (1) to node [] {$\beta$} (3);
\end{tikzpicture}\end{minipage}
        \label{fig:not-normalized}%
    }%
    \subfloat[normalized]{%
        \begin{minipage}[b][2.0cm][t]{2.4cm}\centering\begin{tikzpicture}[node distance=10mm, inner sep=1.5pt, auto, thick] 
    \node[] (0) [] {};
    \node[main] (1) [node distance=7mm,below of=0] {\phantom{$x_i$}};
    \node[] (2) [below left of=1] {};
    \node[] (3) [below right of=1] {};
    \draw[->] (0) to node [] {$\gamma \cdot \normsymb$} (1) ;
    \draw[->,dashed] (1) to node [] {$\alpha / \normsymb$} (2);
    \draw[->] (1) to node [] {$\beta / \normsymb$} (3);
\end{tikzpicture}\end{minipage}
        \label{fig:normalized}%
    }%
    \subfloat[normalization strategies]{%
        \begin{minipage}[b][3.2cm][t]{6.8cm}
        \renewcommand*{\arraystretch}{1.3}
        \renewcommand\cellgape{\Gape[2pt]} 
        \begin{tabular}[b]{l|c|rr}
             & $\normsymb$ & \multicolumn{2}{c}{$\ell^2$-norm $\not\approx 1$} \\\hline\hline
            \textsc{norm-low} & $\alpha$ & 32/389 & (8\%) \\\hline
            \textsc{norm-min} & $\min(\alpha,\beta)$ &  84/416 & (20\%) \\\hline
            \textsc{norm-max} & $\max(\alpha,\beta)$ & 0/384 & (0\%) \\\hline
            \textsc{norm-L2} & \makecell{$(|\alpha|^2 + |\beta|^2) \cdot e^{i\theta}$, \\with $\theta$ s.t. $\alpha/\normsymb \in \mathbb{R}_{+}$} & 0/381 & (0\%) \\
        \end{tabular}%
        \end{minipage}
        \label{tab:norm-strats}
    }
    \centering
    \vspace{-2mm}
    \caption{%
    An arbitrary tuple of edge values $\langle \alpha,\beta,\gamma \rangle$ is not generally in a canonical form, e.g. $\langle 2,6,1 \rangle \equiv \langle 1,3,2 \rangle$. 
    Such a tuple can be normalized by dividing $\alpha$ and $\beta$, and multiplying $\gamma$, by some choice of $\nu$.
    In the example, normalizing both tuples by $\nu = \alpha$ yields $\langle \tfrac{2}{2},\tfrac{6}{2},1\cdot2 \rangle = \langle 1,3,2 \rangle$ and $\langle \tfrac{1}{1},\tfrac{3}{1},2\cdot1 \rangle = \langle 1,3,2 \rangle$.
    The table in \protect\subref{tab:norm-strats} shows different choices for $\nu$, along with an empirical evaluation of the numerical errors when simulating circuits from MQT Bench~\cite{quetschlich2023mqtbench} and checking the $\ell^2$-norm of the output.}
    \label{fig:norm-overview}
\end{figure}

\paragraph{Normalizing edge values}
In order to recognize equivalent nodes, they must be stored in a canonical form. To bring an arbitrary EVDD node into a canonical form, the edge values need to be normalized (\cref{fig:not-normalized,fig:normalized}).
Four different methods, specified in \Cref{tab:norm-strats}, have been implemented. Out of these the first three achieve canonicity by setting one of the edge values to 1, while the fourth is based on the way quantum states are normalized in quantum computing in general. These methods have been tested on the MQT Bench benchmark set~\cite{quetschlich2023mqtbench} (discussed in \cref{sec:experiments}) for circuits up to 30 qubits.
For each run that terminated within the given timeout of 10 minutes we computed the \mbox{$\ell^2$-norm} (the sum of absolute values squared) of the output and checked if this equals 1 ($\pm10^{-3}$) as it should for quantum states. We find that while \textsc{norm-max} and \textsc{norm-L2} did not have any issues, \textsc{norm-low} and \textsc{norm-min} suffered from a significant amount of numerical errors. This can be an indication that having larger values higher up in the decision diagram can increase issues with numerical instability, and it rules out \textsc{norm-low} and \textsc{norm-min} as viable methods.
Between the remaining strategies we find that while \textsc{norm-L2} yields smaller decision diagrams on some instances, \textsc{norm-max} is faster on most instances. 
This difference is likely due to the higher complexity of \textsc{norm-L2}. We therefore choose to set \textsc{norm-max} as the default, and have also used this for the evaluation in \cref{sec:experiments}.  

\paragraph{Parallelization}
Sylvan makes use of the work-stealing framework Lace~\cite{vandijk2014lace}, which can be used for both intra- and inter-operational parallelism with its $\fname{Spawn}$ (fork) and $\fname{Sync}$ (join) commands.  Since we aim to create efficient parallel quantum DD operations, we focus here on intra-operational parallelism, i.e. parallelism within a single (recursive) DD operation. As an example, consider the EVDD algorithm for vector addition, given in \cref{alg:evdd-plus}. The algorithm contains two recursive calls (\cref{line:rec-add0,line:rec-add1}) that are typically executed sequentially. With Lace, these can be parallelized as follows:

{
\setlength{\interspacetitleruled}{0pt}%
\setlength{\algotitleheightrule}{0pt}%
\begin{algorithm}[H]
\setcounter{AlgoLine}{\thelinerecadd-1}


\fname{Spawn}(\add{$\edgeval{A} \cdot \follow{1}{A}$, $\edgeval{B} \cdot \follow{1}{B}$}) \label{line:spawn}
\Comment*[r]{Spawn call as task (fork)} 
$R_0 \gets$ \add{$\edgeval{A} \cdot \follow{0}{A}$, $\edgeval{B} \cdot \follow{0}{B}$} \label{line:call}
\Comment*[r]{Call directly}
$R_1 \gets \fname{Sync}$ \label{line:sync}
\Comment*[r]{Obtain task result (join)}


\end{algorithm}
}

\noindent
The spawned tasks are queued and can be executed by another thread, while the task on \cref{line:call} is executed by the current thread. After finishing its own task, the thread waits for the result of the spawned task.

Since all threads share the same hash table that stores unique nodes, a mechanism is required to protect against race conditions. Sylvan does this using a lock-free node table with atomic compare-and-swap operations, rather than locking (parts) of the table~\cite{vandijk2013multicore}. The same mechanism is also used for the table that stores the edge values (see \cref{alg:find-or-put}).

\section{Use cases}
\label{sec:use-cases}
We now briefly describe the implementation of our two main use cases: simulation and equivalence checking of quantum circuits. Implementations of both are available as command line programs and take as input circuits in the standard quantum circuit format Open QASM 2.0~\cite{cross2017open}, with support for the full set of quantum gates defined by Open QASM's ``qelib1.inc''.
Additionally, Q-Sylvan's C interface can also be used directly, supporting many functions for creating and manipulating vectors and matrices used in quantum computing.\footnote{See \url{https://github.com/System-Verification-Lab/Q-Sylvan\#Documentation}.}

Simulating quantum circuits with DDs is straightforward: first a DD is constructed for the initial all-zero state (i.e. a vector $\begin{\mymatrix} 1 & 0 & 0 & \cdots & 0\end{\mymatrix}^\transpose$), after which for every gate in the circuit the state is updated through matrix-vector multiplication (\cref{alg:evdd-mult}). Q-Sylvan then allows for the final state to be either output directly, or to draw samples from it through simulated quantum measurements.




Our second use case is quantum circuit equivalence checking, which is defined as follows: given two $n$-qubit quantum circuits $U = \{U_1, \dots, U_m\}$ and $V = \{V_1, \dots, V_\ell \}$, where $U_i$ and $V_i$ are the individual gates composing the circuits, are $U$ and $V$ represented by the same matrices up to a global factor? I.e. does there exist some $c \in \mathbb{C}$ such that $U = cV$ (or equivalently we write $U \equiv V$)?
There are a variety of ways to check quantum circuit equivalence. The naive method is to compute the full $2^n \times 2^n$ sized matrices $U$ and $V$ through multiplication of their individual gates. However, much like computing the composition of transition relations tends to be inefficient with DDs in classical model checking~\cite{matsunaga1993computing}, the DDs resulting from multiplying quantum gates together tend to be much larger than those obtained from only updating states. However, as discussed below, one can be clever in choosing the order in which the gates are multiplied.


We implement two equivalence checking algorithms: ``alternating'' and ``Pauli''.
The alternating algorithm was proposed in~\cite{burgholzer2020advanced}, and the idea is to first rewrite $U \equiv V$ as $UV^\dagger \equiv I$, where $I$ is the identity matrix, $U V^\dagger =  U_m \dots U_2 U_1 V^\dagger_1 V^\dagger_2 \dots V^\dagger_\ell$, and $V_i^\dagger = \left(V_i^*\right)^\transpose$ is the conjugate transpose of $V_i$. This product can then be computed from the inside out, e.g. if $m = \ell$ the computation order would be $(U_m \dots (U_2 (U_1 V^\dagger_1) V^\dagger_2) \dots V^\dagger_\ell)$. If (w.l.o.g.) $m \geq \ell$ the algorithm takes $\tfrac{m}{\ell}$ gates from $U$ for every gate from $V$. The motivation for this approach is that when $U$ and $V$ are identical, every step of the computation yields the identity matrix, and thus the computation remains easy. When $U$ and $V$ are equivalent but not identical, this approach can still heuristically yield matrices that have small DD encodings.
The Pauli algorithm is based on~\cite[Thm.1]{thanos2023fast}, which notes that $U \equiv V \iff \forall j \in \{0, \dots, n-1\} (U X_j U^\dagger = V X_j V^\dagger) \land (U Z_j U^\dagger = V Z_j V^\dagger)$, where $X_j$ and $Z_j$ special matrices (specifically tensor products of identity and Pauli matrices) that have an efficient classical description. Similar to the alternating algorithm, the terms can be computed from the inside out, e.g. compute $U X_j U^\dagger$ as $(U_m \dots (U_2 (U_1 X_j U_1^\dagger) U_2^\dagger) \dots U_m^\dagger)$. The motivation here is that when $U$ and $V$ consist of a particular subset of quantum gates called Clifford gates this computation is provably efficient (i.e. polynomial time), and when they consist of more general gates there can still be heuristic benefits from the compression DDs provide. Ours is the first DD-based implementation of this algorithm.




\section{Empirical evaluation}
\label{sec:experiments}
We evaluate Q-Sylvan against several state-of-the-art tools.\footnote{Reproducible benchmarks are available at \url{https://github.com/sebastiaanbrand/q-sylvan-benchmarks}.}
The single and 8-core results were obtained on an AMD Ryzen 7 5800x CPU with 8 cores and 64 GB of memory.
The \mbox{64-core} results were obtained on a machine with two AMD EPYC 7601 CPUs with 32 physical cores each (64 in total) and 1 TB of memory. 

\subsection{Simulation}

For testing simulation performance we use two sets of quantum circuits in the Open QASM~2.0 format. The first, MQT Bench~\cite{quetschlich2023mqtbench}, contains 22 types of quantum circuits for which the number of qubits can be arbitrarily scaled, as well as 6 types of non-scalable circuits with varying numbers of qubits. To obtain a greater variety of circuits, the second dataset we include is one generated by KetGPT~\cite{apak2024ketgpt}, an instance of ChatGPT that was trained on the MQT Bench dataset, and consists of 1000 quantum circuits with varying numbers of qubits.
The reported runtimes only include computing the final state vector. Simulating measurements given this state vector constitutes negligible overhead~\cite{zulehner2018advanced}.

We first test Q-Sylvan's single-core performance against the well-established EVDD-based quantum circuit simulator MQT DDSIM~\cite{zulehner2018advanced}, shown in \Cref{fig:qsylvan_vs_mqt}.
We find that while DDSIM outperforms Q-Sylvan on the smaller instances (presumably in part due to a more efficient initialization), on larger instances (where either tool takes $\geq 10$ seconds) Q-Sylvan outperforms DDSIM on 61\% of the MQT Bench circuits, and 30\% of KetGPT circuits. A comparison against Quasimodo~\cite{sistla2023symbolic} 
(\arxiv{\cref{app:additional-res}, \Cref{tab:evdd-cflobdd-simulation}}{see~\cite[App. A]{arxivversion}}) 
shows that, while CFLOBDDs can theoretically achieve much greater compactness than EVDDs, on this benchmark set they beat EVDDs only on the GHZ circuit. We omit SliQSim~\cite{tsai2021bitslicing}, as it does not support this benchmark set due to a lack of rotation-gate support.

{
\captionsetup[subfloat]{captionskip=-1.5mm}
\begin{figure}[t]
    \centering
    \subfloat[MQT Bench (10--50 qubits)]{%
        \includegraphics[width=0.5\textwidth]{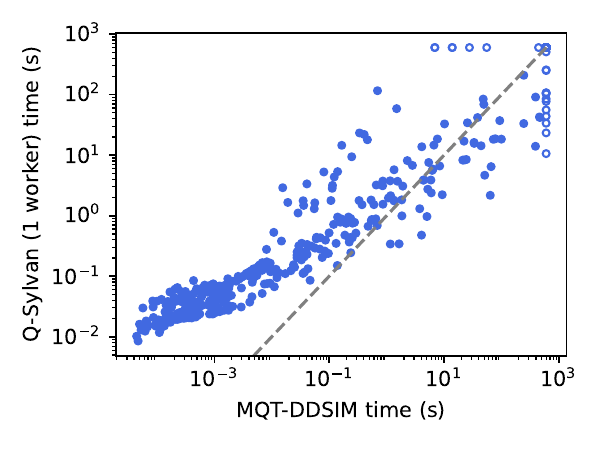}%
    }%
    \subfloat[KetGPT]{%
        \includegraphics[width=0.5\textwidth]{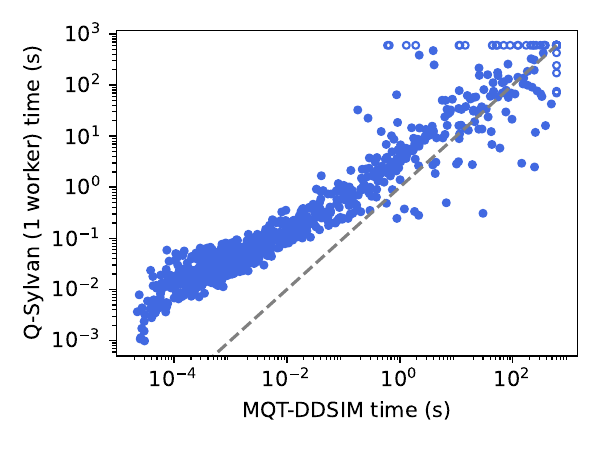}%
    }%
    \vspace{-4mm}
    \caption{Q-Sylvan vs DDSIM. Both order variables according to the qubit ordering in the QASM file. Open markers indicate timeouts. For both plots, we verified the full state vector output of both tools up to 20 qubits. While DDSIM is faster on the smaller circuits, on circuits where either tool takes $\geq 10$ seconds \mbox{Q-Sylvan} beats DDSIM in 61\% of MQT Bench circuits, and 30\% of KetGPT circuits.}
    \label{fig:qsylvan_vs_mqt}
    \vspace{-2mm}
\end{figure}
}

\begin{figure}[p]
    \centering
    \subfloat[Speedup percentiles]{%
        \scalebox{1}{%
        \setlength{\tabcolsep}{0.4em}
\def\arraystretch{1}
\begin{tabular}{c|c||ccc|ccc}
    \multirow{2}{*}{dataset} & \multirow{2}{*}{sharing} & \multicolumn{3}{c|}{8-core speedup} & \multicolumn{3}{c}{64-core speedup}\\
     & & $P_{90}$ & $P_{95}$ & $P_{99}$ 
       & $P_{90}$ & $P_{95}$ & $P_{99}$ \\\hline
     \multirow{3}{*}{\begin{tabular}{c}MQT Bench\end{tabular}}
     & high sharing & $\times 1.1$ & $\times 1.3$ & $\times 1.7$
                    & $\times 0.5$ & $\times 0.5$ & $\times 0.7$ \\
     & some sharing & $\times 3.6$ & $\times 4.2$ & $\times 5.8$
                    & $\times 1.2$ & $\times 2.9$ & $\times 4.2$ \\
     & no sharing   & $\times 6.0$ & $\times 6.2$ & $\times 6.4$
                    & $\times 11$  & $\times 13$  & $\times 14$ \\\hline
     \multirow{3}{*}{KetGPT}
     & high sharing & $\times 1.4$ & $\times 1.9$ & $\times 2.7$
                    & $\times 0.7$ & $\times 0.9$ & $\times 1.2$ \\
     & some sharing & $\times 5.9$ & $\times 6.4$ & $\times 7.2$
                    & $\times 8.3$ & $\times 11$  & $\times 18$  \\
     & no sharing   & $\times 3.5$ & $\times 4.5$ & $\times 4.6$
                    & $\times 1.0$ & $\times 2.1$ & $\times 2.6$ \\
\end{tabular}

        }
        \label{tab:speedups}
    }\\
    \captionsetup[subfloat]{farskip=.05cm,captionskip=-.2cm}
    \subfloat[DD sizes on MQT Bench]{%
        \includegraphics[width=0.5\textwidth]{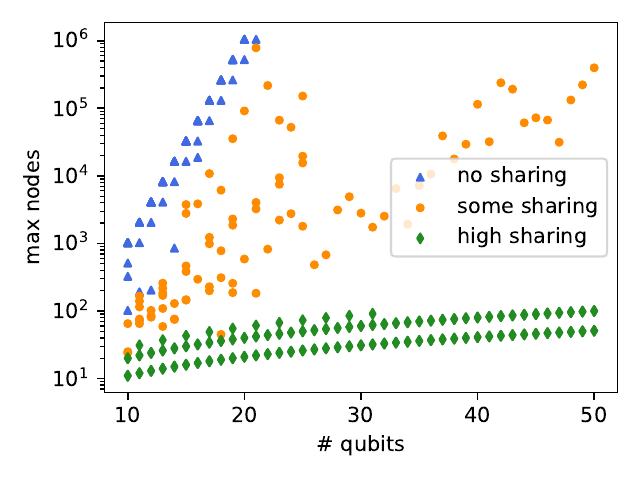}%
        \label{fig:ddsizes-mqtbench}%
    }
    \subfloat[DD sizes on KetGPT]{%
        \includegraphics[width=0.5\textwidth]{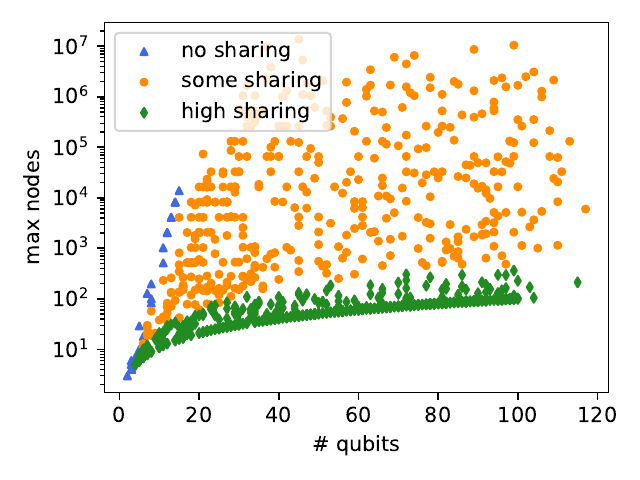}%
        \label{fig:ddsizes-ketgpt}%
    }\\
    \subfloat[8-core runtime on MQT Bench]{%
        \includegraphics[width=0.5\textwidth]{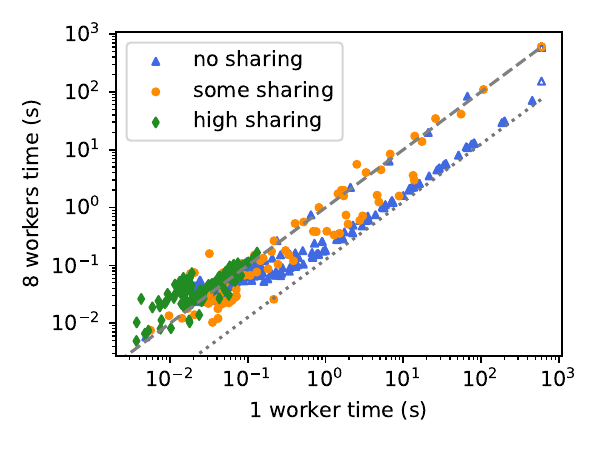}%
        \label{fig:8core-mqtbench}%
    }
    \subfloat[8-core runtime on KetGPT]{%
        \includegraphics[width=0.5\textwidth]{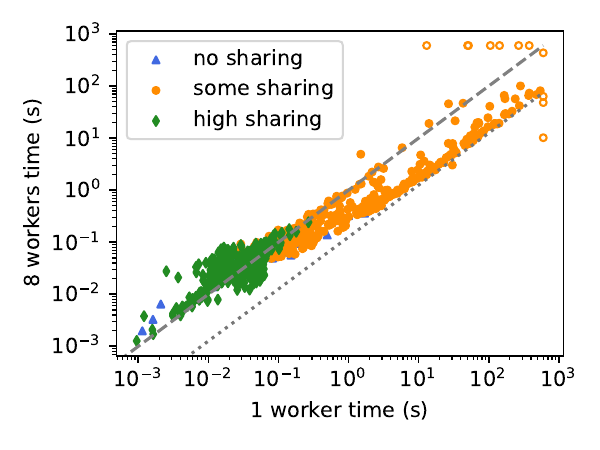}%
        \label{fig:8core-ketgpt}%
    }\\
    \subfloat[64-core runtime on MQT Bench]{%
        \includegraphics[width=0.5\textwidth]{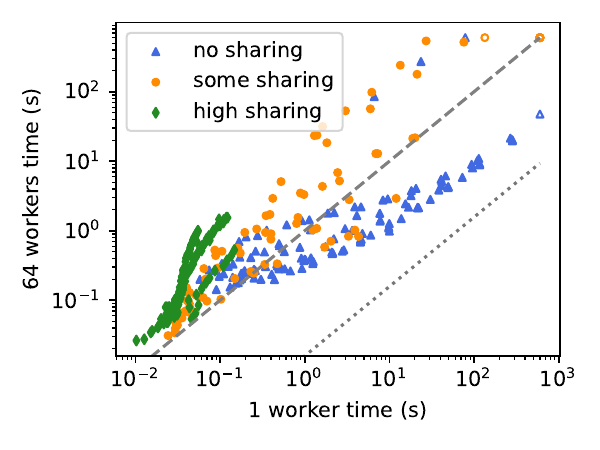}%
        \label{fig:64core-mqtbench}%
    }
    \subfloat[64-core runtime on KetGPT]{%
        \includegraphics[width=0.5\textwidth]{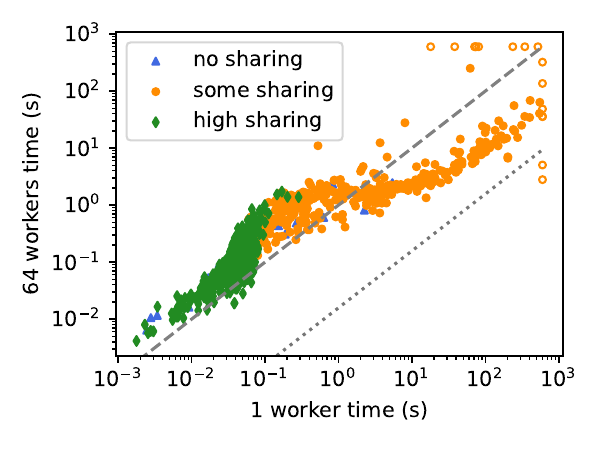}%
        \label{fig:64core-ketgpt}%
    }\\
    \caption{Parallel performance. Open markers indicate timeouts, dashed lines indicate equal performance, and dotted lines indicate a $\times k$ speedup for $k$-cores.}
    \label{fig:qsylvan_parallel}
\end{figure}

Second, we evaluate Q-Sylvan's parallel performance. Since, when comparing single-core performance, Q-Sylvan performs well on a significant fraction of larger circuits, we evaluate Q-Sylvan's multi-core performance against its own single-core performance.
As noted in~\cite{hillmich2020concurrency}, parallel speedups can greatly depend on the compactness of the decision diagrams. For example, it is easier to obtain speedups on a DD that is effectively a binary tree than it is to obtain speedups on a very compact DD. To evaluate this effect, we split the DDs resulting from the different benchmarks into three categories: \emph{no sharing} (DDs which, for $n$ qubits, have almost $2^n$ nodes), \emph{high sharing} (DDs with less than $n \log n$ nodes), and \emph{some sharing} (the DDs which fall in between these categories). \Cref{fig:ddsizes-mqtbench,fig:ddsizes-ketgpt} show the distribution of decision diagram sizes for both benchmark sets. 
Obtaining speedups on the ``no sharing'' category would be the easiest, however these are instances where EVDDs offer little or no benefit. On the other hand, obtaining speedups on the ``high sharing'' category would be very difficult, since the DDs are so compact that little work remains that can be done in parallel. We are therefore mostly interested in speedups in the ``some sharing'' category. 

The full parallel performance results are shown in \Cref{fig:8core-mqtbench,fig:8core-ketgpt,fig:64core-mqtbench,fig:64core-ketgpt}, and a summary of the speedups at different percentiles is given in \Cref{tab:speedups}.
We find that, on the ``some sharing'' instances, Q-Sylvan is able to obtain speedups of up to $\times 7.2$ and $\times 18$ for 8 and 64 cores respectively. 
Unfortunately, we are unable to test directly against the two other parallel EVDD implementations~\cite{hillmich2020concurrency,li2024parallelizing}, as one~\cite{hillmich2020concurrency} does not have an implementation available anymore (neither public nor private), while the other~\cite{li2024parallelizing} hard codes two circuits but does not allow for the parallel simulation of arbitrary circuits.
However, for comparison, the first~\cite{hillmich2020concurrency} reported speedups of up to $\times 3$ using 32 cores on ``no sharing'' circuits, while the second~\cite{li2024parallelizing} reported speedups of $\times 2\textrm{--}3$ using 16 cores on ``some sharing'' circuits.

On the MQT Bench dataset we find that while the 8 core speedups are promising, the 64 core results are less strong. This could in part be caused by a higher communication overhead, since the 64 cores are split between two separate CPUs. Additionally, it might be that larger benchmarks (not just in numbers of qubits, but in number of DD nodes) are required to show greater speedups for 64 cores.
The underperformance of Q-Sylvan on the ``no sharing'' subset of the KetGPT circuits can be explained by the fact that those types of circuits are almost not present in the KetGPT dataset.


\subsection{Equivalence checking}
Next we compare our two equivalence checking implementations against two recent tools: MQT QCEC~\cite{burgholzer2020advanced} and Quokka-Sharp~\cite{mei2024equivalence}. We omit SliQEC~\cite{wei2022accurate} as it supports very few circuits due to a lack of rotation-gate support. MQT QCEC is a portfolio of different algorithms and makes use of both EVDDs as well a ZX-calculus, while Quokka-Sharp is based on weighted model counting. Both support multi-core computations.
We use the benchmark set from~\cite{mei2024equivalence}, consisting of 78 pairs of equivalent circuits, and 146 non-equivalent pairs.

An overview of the number of completed benchmarks, as well as the speedups for 8-core computations, is given in \cref{tab:eqcheck-summary}. 
\arxiv{Detailed single-core results are given in \cref{tab:eqcheck-full1,tab:eqcheck-full2,tab:noneqcheck-full1,tab:noneqcheck-full2} in \Cref{app:additional-res}.}{Detailed single-core results can be found in~\cite[App. A]{arxivversion}.}
We find that Q-Sylvan shows a similar potential for speedups on the task of equivalence checking as it does on simulation. Using 8 cores it achieves a speedup of $\times 5.8$ on both the equivalent and non-equivalent sets of circuits.
Although \mbox{Q-Sylvan} is unable to match the single-core performance of QCEC, it still manages to solve some instances (19\% of equivalent and 5\% of non-equivalent circuits) faster than QCEC.
QCEC, being a portfolio method, can parallelize computations naively, and greatly benefits from this on the equivalent instances. Its greater than $\times 8$ parallel performance on these instances is likely a result of the way it prioritizes testing for non-equivalence over verifying equivalence when running on a single core. This might also explain its decrease in the number of solved non-equivalent instances when using more cores.

Overall we find that \mbox{Q-Sylvan's} equivalence checking algorithms show promising parallel performance, and would likely benefit from being embedded into a portfolio approach.

\begin{table}[t]
    \centering
    \caption{Comparison of Q-Sylvan, Quokka-Sharp, and MQT QCEC, separated into equivalent and non-equivalent benchmarks. Timeout set to 5 minutes. 
    }
    \label{tab:eqcheck-summary}
    \vspace{-2mm}
    \setlength{\tabcolsep}{0.3em}%
    \def\arraystretch{1.0}%
    \scalebox{1}{
    \begin{tabular}{l|r|cc|cc|cc|cc}
        \multicolumn{2}{c|}{} 
         & \multicolumn{2}{l|}{\multirow{2}{*}{\shortstack{Q-Sylvan \\ alternating}}}
         & \multicolumn{2}{l|}{\multirow{2}{*}{\shortstack{Q-Sylvan \\ Pauli}}}
         & \multicolumn{2}{l|}{\multirow{2}{*}{Quokka-Sharp}}
         & \multicolumn{2}{l}{\multirow{2}{*}{MQT QCEC}} \\ 
         \multicolumn{2}{c|}{} & & & & & & & & \\ \hline
         \multicolumn{2}{c|}{cores} & 1 & 8 & 1 & 8 & 1 & 8 & 1 & 8 \\ \hline\hline
       \multirow{2}{*}{equiv} 
        & \% completed & 59\% & 62\%  & 49\% & 59\% & 39\% & 39\% & 64\% & 67\%  \\
        & runtime reduction &  
        & $\times 5.8$  &  & $\times 2.1$ &  & $\times 2.2$ &  & $\times 15$ \\\hline
       \multirow{2}{*}{non-equiv} 
        & \% completed & 56\% & 60\%  & 47\% & 58\% & 50\% & 56\% & 82\% & 69\% \\
        & runtime reduction &  
        & $\times 5.8$  &  & $\times 2.2$ &  & $\times 2.1$ &  & $\times 1.3$ \\\hline
    \end{tabular}%
    }
\end{table}


\subsubsection*{Acknowledgements}
This work was supported by the NEASQC project, funded by the European Union’s Horizon 2020, Grant Agreement No. 951821.

%
%
%
\bibliographystyle{splncs04}
\bibliography{references}

\arxiv{
\appendix
\section{Additional results}
\label{app:additional-res}

\begin{table}[H]
    \setlength{\tabcolsep}{0.7em}
    \def\arraystretch{1}
    \centering
    \caption{Simulation times on a subset of MQT Bench circuits.}
    \label{tab:evdd-cflobdd-simulation}
    \begin{tabular}{lr||rrr}
Algorithm & $n$ & \shortstack{Q-Sylvan \\ (EVDD)} & \shortstack{MQT DDSIM \\(EVDD)} & \shortstack{Quasimodo \\ (CLFOBDD)} \\
\hline
Amplitude estimation & 8 & 0.09 & \textbf{0.01} & 0.80 \\
Amplitude estimation & 16 & \textbf{8.06} & 19.61 & > 600 \\
Amplitude estimation & 32 & - & > 600 & > 600 \\
Deutsch-Jozsa & 8 & 0.03 & \textbf{0.00} & 0.21 \\
Deutsch-Jozsa & 16 & 0.03 & \textbf{0.00} & > 600 \\
Deutsch-Jozsa & 32 & 0.04 & \textbf{0.00} & > 600 \\
GHZ & 8 & 0.01 & \textbf{0.00} & 0.02 \\
GHZ & 16 & 0.03 & \textbf{0.00} & 0.02 \\
GHZ & 32 & 0.04 & \textbf{0.00} & 0.02 \\
Graphstate & 8 & 0.04 & \textbf{0.00} & 0.13 \\
Graphstate & 16 & 0.05 & \textbf{0.00} & > 600 \\
Graphstate & 32 & 0.19 & \textbf{0.01} & > 600 \\
QFT & 8 & 0.06 & \textbf{0.00} & 0.06 \\
QFT & 16 & 0.14 & \textbf{0.00} & > 600 \\
QFT & 32 & 0.54 & \textbf{0.02} & > 600 \\
QFT entangled & 8 & 0.09 & \textbf{0.00} & 0.13 \\
QFT entangled & 16 & \textbf{8.23} & 186.47 & > 600 \\
QFT entangled & 32 & > 600 & > 600 & > 600 \\
QNN & 8 & 0.12 & \textbf{0.01} & 0.76 \\
QNN & 16 & 8.77 & \textbf{2.95} & > 600 \\
QNN & 32 & - & > 600 & > 600 \\
QPE exact & 8 & 0.06 & \textbf{0.00} & 0.17 \\
QPE exact & 16 & 0.65 & \textbf{0.02} & > 600 \\
QPE exact & 32 & - & > 600 & > 600 \\
QPE inexact & 8 & 0.07 & \textbf{0.00} & 0.17 \\
QPE inexact & 16 & \textbf{0.97} & 2.75 & > 600 \\
QPE inexact & 32 & - & > 600 & > 600 \\
Random & 8 & 0.17 & \textbf{0.01} & 0.75 \\
Random & 16 & 70.66 & \textbf{28.18} & > 600 \\
Random & 32 & - & > 600 & > 600 \\
Real amp. random & 8 & 0.11 & \textbf{0.01} & 0.47 \\
Real amp. random & 16 & \textbf{16.53} & 37.25 & > 600 \\
Real amp. random & 32 & - & > 600 & > 600 \\
SU2 random & 8 & 0.13 & \textbf{0.01} & 0.48 \\
SU2 random & 16 & \textbf{17.08} & 39.88 & > 600 \\
SU2 random & 32 & - & > 600 & > 600 \\
Two-local random & 8 & 0.12 & \textbf{0.01} & 0.47 \\
Two-local random & 16 & \textbf{16.97} & 37.41 & > 600 \\
Two-local random & 32 & - & > 600 & > 600 \\
W-state & 8 & 0.06 & \textbf{0.00} & 0.12 \\
W-state & 16 & 0.06 & \textbf{0.00} & > 600 \\
W-state & 32 & 0.12 & \textbf{0.00} & > 600 \\
\end{tabular}

\end{table}


\begin{table}[H]
    \centering
    \caption{Detailed single-core equivalence checking results. ``-'' indicates an out-of-memory termination, and ``$\times$'' indicates an incorrect result.}
    \label{tab:eqcheck-full1}
    \scalebox{0.8}{
    \setlength{\tabcolsep}{0.4em}
\def\arraystretch{1}
\begin{tabular}{l||rrr||rrrr}
\multirow{2}{*}{Algorithm} & 
\multirow{2}{*}{$n$} & 
\multirow{2}{*}{$|G|$} & 
\multirow{2}{*}{$|G'|$} & 
\multirow{2}{*}{\shortstack{Q-Sylvan \\ alternating}} & 
\multirow{2}{*}{\shortstack{Q-Sylvan \\ Pauli}} & 
\multirow{2}{*}{Quokka-Sharp} & 
\multirow{2}{*}{MQT QCEC} \\
& & & & & & & \\
\hline
Amplitude estimation & 16 & 830 & 802 & - & - & > 300 & > 300 \\
Amplitude estimation & 32 & 2950 & 2862 & > 300 & - & > 300 & > 300 \\
Amplitude estimation & 64 & 11030 & 7728 & > 300 & - & > 300 & > 300 \\
Deutsch-Jozsa & 16 & 127 & 67 & 0.04 & 1.44 & 0.11 & \textbf{0.02} \\
Deutsch-Jozsa & 32 & 249 & 129 & 0.11 & 7.18 & 0.29 & \textbf{0.04} \\
Deutsch-Jozsa & 64 & 507 & 259 & 0.31 & 52.12 & 1.05 & \textbf{0.12} \\
GHZ & 16 & 18 & 46 & \textbf{0.01} & 0.22 & 0.05 & 0.01 \\
GHZ & 32 & 34 & 94 & \textbf{0.02} & 1.23 & 0.14 & 0.02 \\
GHZ & 64 & 66 & 190 & \textbf{0.05} & 7.94 & 0.42 & 0.07 \\
Graphstate & 16 & 160 & 32 & 0.05 & 2.20 & 0.12 & \textbf{0.02} \\
Graphstate & 32 & 320 & 64 & \textbf{0.20} & 9.14 & 0.48 & 0.32 \\
Graphstate & 64 & 640 & 128 & 4.76 & 56.83 & \textbf{2.44} & > 300 \\
Groundstate & 4 & 180 & 36 & 0.03 & 0.52 & 1.92 & \textbf{0.01} \\
Groundstate & 12 & 1212 & 164 & \textbf{0.31} & 56.80 & > 300 & 1.51 \\
Groundstate & 14 & 1610 & 206 & \textbf{0.44} & 270.09 & > 300 & 14.84 \\
Grover (no ancilla) & 5 & 499 & 629 & 0.26 & 15.51 & $\times$ & \textbf{0.06} \\
Grover (no ancilla) & 6 & 1568 & 1870 & 13.92 & 210.11 & > 300 & \textbf{0.48} \\
Grover (no ancilla) & 7 & 3751 & 5783 & \textbf{157.44} & > 300 & > 300 & > 300 \\
Grover (v-chain) & 5 & 529 & 632 & 0.28 & 18.37 & $\times$ & \textbf{0.06} \\
Grover (v-chain) & 7 & 1224 & 1627 & 37.87 & > 300 & > 300 & \textbf{27.02} \\
Grover (v-chain) & 9 & 3187 & 4815 & > 300 & > 300 & > 300 & > 300 \\
Portfolio QAOA & 5 & 195 & 236 & 0.12 & 6.15 & 128.48 & \textbf{0.04} \\
Portfolio QAOA & 6 & 261 & 356 & 1.78 & 36.35 & > 300 & \textbf{0.12} \\
Portfolio QAOA & 7 & 336 & 481 & 8.72 & 204.75 & > 300 & \textbf{0.59} \\
Portfolio VQE & 5 & 310 & 131 & 0.12 & 5.34 & 190.61 & \textbf{0.03} \\
Portfolio VQE & 6 & 435 & 151 & 1.20 & 27.99 & > 300 & \textbf{0.06} \\
Portfolio VQE & 7 & 581 & 218 & 0.35 & 147.63 & > 300 & \textbf{0.13} \\
Pricing call & 5 & 240 & 166 & 0.09 & 1.23 & 8.12 & \textbf{0.02} \\
Pricing call & 7 & 422 & 277 & 0.20 & 20.07 & > 300 & \textbf{0.10} \\
Pricing call & 9 & 624 & 396 & 233.38 & > 300 & > 300 & \textbf{0.41} \\
Pricing put & 5 & 240 & 192 & 0.10 & 1.79 & 8.69 & \textbf{0.02} \\
Pricing put & 7 & 432 & 297 & 3.95 & 17.88 & > 300 & \textbf{0.11} \\
Pricing put & 9 & 654 & 428 & 265.13 & > 300 & > 300 & \textbf{0.64} \\
QAOA & 7 & 133 & 117 & 0.12 & 1.06 & 0.59 & \textbf{0.03} \\
QAOA & 9 & 171 & 296 & 1.29 & 2.71 & 1.34 & \textbf{0.33} \\
QAOA & 11 & 209 & 359 & > 300 & 4.12 & 1.52 & \textbf{0.86} \\
QFT & 2 & 14 & 14 & \textbf{0.01} & 0.02 & 0.01 & 0.01 \\
QFT & 8 & 176 & 228 & 1.16 & 10.59 & 24.53 & \textbf{0.05} \\
QFT & 16 & 672 & 814 & > 300 & > 300 & > 300 & \textbf{10.34} \\
QFT entangled & 16 & 690 & 853 & > 300 & > 300 & > 300 & > 300 \\
QFT entangled & 32 & 2658 & 3068 & - & - & > 300 & > 300 \\
QFT entangled & 64 & 10434 & 9387 & > 300 & > 300 & > 300 & > 300 \\
QNN & 2 & 43 & 36 & 0.01 & 0.09 & 0.09 & \textbf{0.01} \\
QNN & 8 & 319 & 494 & 12.14 & > 300 & > 300 & \textbf{0.37} \\
QNN & 16 & 1023 & 2002 & > 300 & > 300 & > 300 & > 300 \\
QPE exact & 16 & 712 & 837 & > 300 & > 300 & 300.00 & > 300 \\
QPE exact & 32 & 2712 & 3276 & > 300 & > 300 & > 300 & > 300 \\
QPE exact & 64 & 10552 & 9680 & > 300 & > 300 & > 300 & > 300 \\
QPE inexact & 16 & 712 & 848 & > 300 & > 300 & > 300 & \textbf{229.12} \\
QPE inexact & 32 & 2712 & 3179 & > 300 & > 300 & > 300 & > 300 \\
QPE inexact & 64 & 10552 & 9695 & > 300 & > 300 & > 300 & > 300 \\
\end{tabular}

    }
\end{table}

\begin{table}[H]
    \centering
    \caption{Continued from \Cref{tab:eqcheck-full1}.}
    \label{tab:eqcheck-full2}
    \scalebox{0.8}{
    \setlength{\tabcolsep}{0.4em}
\def\arraystretch{1}
\begin{tabular}{l||rrr||rrrr}
\multirow{2}{*}{Algorithm} & 
\multirow{2}{*}{$n$} & 
\multirow{2}{*}{$|G|$} & 
\multirow{2}{*}{$|G'|$} & 
\multirow{2}{*}{\shortstack{Q-Sylvan \\ alternating}} & 
\multirow{2}{*}{\shortstack{Q-Sylvan \\ Pauli}} & 
\multirow{2}{*}{Quokka-Sharp} & 
\multirow{2}{*}{MQT QCEC} \\
& & & & & & & \\
\hline
Q-walk (no ancilla) & 6 & 2457 & 3003 & 17.89 & > 300 & > 300 & \textbf{0.83} \\
Q-walk (no ancilla) & 7 & 4761 & 6231 & 216.96 & > 300 & > 300 & \textbf{2.06} \\
Q-walk (no ancilla) & 8 & 9369 & 12486 & > 300 & > 300 & > 300 & > 300 \\
Q-walk (v-chain) & 5 & 417 & 398 & 0.21 & 2.42 & $\times$ & \textbf{0.03} \\
Q-walk (v-chain) & 7 & 849 & 840 & 0.87 & > 300 & > 300 & \textbf{0.21} \\
Q-walk (v-chain) & 9 & 1437 & 1500 & 3.78 & > 300 & > 300 & \textbf{1.56} \\
Real amp. random & 16 & 680 & 679 & > 300 & > 300 & > 300 & > 300 \\
Real amp. random & 32 & 2128 & 2215 & > 300 & > 300 & > 300 & > 300 \\
Real amp. random & 64 & 7328 & 7411 & > 300 & > 300 & > 300 & > 300 \\
Routing & 2 & 43 & 29 & 0.02 & 0.09 & 0.09 & \textbf{0.01} \\
Routing & 6 & 135 & 142 & 0.06 & 6.82 & 135.12 & \textbf{0.03} \\
Routing & 12 & 273 & 409 & - & > 300 & > 300 & \textbf{4.79} \\
SU2 random & 16 & 744 & 378 & - & > 300 & > 300 & > 300 \\
SU2 random & 32 & 2256 & 762 & > 300 & > 300 & > 300 & > 300 \\
SU2 random & 64 & 7584 & 1530 & > 300 & > 300 & > 300 & > 300 \\
TSP & 4 & 225 & 86 & 0.07 & 1.65 & 18.66 & \textbf{0.01} \\
TSP & 9 & 550 & 315 & 257.19 & > 300 & > 300 & \textbf{0.91} \\
TSP & 16 & 1005 & 623 & > 300 & > 300 & > 300 & > 300 \\
Two-local random & 16 & 680 & 679 & > 300 & > 300 & > 300 & > 300 \\
Two-local random & 32 & 2128 & 2215 & > 300 & > 300 & > 300 & > 300 \\
Two-local random & 64 & 7328 & 7411 & > 300 & > 300 & > 300 & > 300 \\
VQE & 5 & 83 & 83 & 0.03 & 0.73 & 1.28 & \textbf{0.02} \\
VQE & 10 & 168 & 221 & 144.14 & > 300 & > 300 & \textbf{2.35} \\
VQE & 15 & 253 & 349 & > 300 & - & > 300 & > 300 \\
W-state & 16 & 271 & 242 & > 300 & 5.82 & 1.72 & \textbf{0.04} \\
W-state & 32 & 559 & 498 & > 300 & 32.29 & 8.66 & \textbf{0.76} \\
W-state & 64 & 1135 & 1010 & > 300 & > 300 & \textbf{54.25} & > 300 \\
\end{tabular}

    }
    \vspace{1mm}
\end{table}
\vspace{-2em}

\begin{table}[H]
    \centering
    \caption{Detailed single-core non-equivalence checking results. `-'' indicates an out-of-memory termination.}
    \label{tab:noneqcheck-full1}
    \scalebox{0.8}{
    \begin{tabular}{l||rrr||rrrr||rrrr}
&  &  &  & \multicolumn{4}{c||}{1 gate missing} & \multicolumn{4}{c}{Flipped control/target of 1 gate} \\
\multirow{2}{*}{Algorithm} & 
\multirow{2}{*}{$n$} & 
\multirow{2}{*}{$|G|$} & 
\multirow{2}{*}{$|G'|$} & 
\multirow{2}{*}{\shortstack{Q-Sylvan \\ alter.}} & 
\multirow{2}{*}{\shortstack{Q-Sylvan \\ Pauli}} & 
\multirow{2}{*}{\shortstack{Quokka \\Sharp}} & 
\multirow{2}{*}{\shortstack{MQT \\QCEC}} & 
\multirow{2}{*}{\shortstack{Q-Sylvan \\ alter.}} & 
\multirow{2}{*}{\shortstack{Q-Sylvan \\ Pauli}} & 
\multirow{2}{*}{\shortstack{Quokka \\Sharp}} & 
\multirow{2}{*}{\shortstack{MQT \\QCEC}} \\
& & & & & & & & & & & \\
\hline
Amplitude est. & 16 & 830 & 802 & - & > 300 & > 300 & \textbf{60.59} & - & - & > 300 & \textbf{44.26} \\
Amplitude est. & 32 & 2950 & 2862 & > 300 & - & > 300 & > 300 & > 300 & - & > 300 & > 300 \\
Amplitude est. & 64 & 11030 & 7728 & > 300 & - & > 300 & > 300 & > 300 & - & > 300 & > 300 \\
Deutsch-Jozsa & 16 & 127 & 67 & 0.04 & 1.44 & 0.07 & \textbf{0.01} & 0.04 & 1.71 & 0.08 & \textbf{0.01} \\
Deutsch-Jozsa & 32 & 249 & 129 & 0.13 & 7.76 & 0.17 & \textbf{0.01} & 0.12 & 7.09 & 0.16 & \textbf{0.01} \\
Deutsch-Jozsa & 64 & 507 & 259 & 0.33 & 51.11 & 0.63 & \textbf{0.01} & 0.30 & 48.49 & 0.79 & \textbf{0.02} \\
GHZ & 16 & 18 & 46 & 0.01 & 0.22 & 0.03 & \textbf{0.00} &  &   &   &   \\
GHZ & 32 & 34 & 94 & 0.01 & 1.24 & 0.10 & \textbf{0.01} &  &   &   &   \\
GHZ & 64 & 66 & 190 & 0.05 & 7.88 & 0.31 & \textbf{0.01} &  &   &   &   \\
Graphstate & 16 & 160 & 32 & 0.05 & 1.84 & 0.03 & \textbf{0.01} &   &   &   &   \\
Graphstate & 32 & 320 & 64 & 0.20 & 9.02 & 0.14 & \textbf{0.02} &   &   &   &   \\
Graphstate & 64 & 640 & 128 & 4.64 & 58.70 & \textbf{0.63} & 96.43 &   &   &   &   \\
Groundstate & 4 & 180 & 36 & 0.04 & 0.65 & 0.38 & \textbf{0.00} &   &   &   &   \\
Groundstate & 12 & 1212 & 164 & 0.38 & 55.73 & > 300 & \textbf{0.10} & 0.35 & 50.30 & > 300 & \textbf{0.12} \\
Groundstate & 14 & 1610 & 206 & \textbf{0.53} & 278.44 & > 300 & 0.85 & \textbf{0.42} & 137.66 & > 300 & 0.74 \\
Grover (no an.) & 5 & 499 & 629 & 0.42 & 15.81 & 33.98 & \textbf{0.01} & 0.29 & 14.78 & 36.81 & \textbf{0.01} \\
Grover (no an.) & 6 & 1568 & 1870 & 14.19 & 226.17 & > 300 & \textbf{0.02} & 13.40 & 163.46 & > 300 & \textbf{0.03} \\
Grover (no an.) & 7 & 3751 & 5783 & 169.09 & > 300 & > 300 & \textbf{0.15} & 149.53 & > 300 & > 300 & \textbf{0.14} \\
Grover (v-chain) & 5 & 529 & 632 & 0.35 & 21.76 & 89.83 & \textbf{0.00} & 0.28 & 19.09 & 90.73 & \textbf{0.01} \\
Grover (v-chain) & 7 & 1224 & 1627 & 16.72 & > 300 & > 300 & \textbf{0.04} & 37.19 & > 300 & > 300 & \textbf{0.01} \\
Grover (v-chain) & 9 & 3187 & 4815 & > 300 & > 300 & > 300 & \textbf{0.38} & > 300 & > 300 & > 300 & \textbf{0.44} \\
\end{tabular}

    }
\end{table}

\begin{table}[H]
    \centering
    \caption{Continued from \Cref{tab:noneqcheck-full1}.}
    \label{tab:noneqcheck-full2}
    \scalebox{0.8}{
    \begin{tabular}{l||rrr||rrrr||rrrr}
&  &  &  & \multicolumn{4}{c||}{1 gate missing} & \multicolumn{4}{c}{Flipped control/target of 1 gate} \\
\multirow{2}{*}{Algorithm} & 
\multirow{2}{*}{$n$} & 
\multirow{2}{*}{$|G|$} & 
\multirow{2}{*}{$|G'|$} & 
\multirow{2}{*}{\shortstack{Q-Sylvan \\ alter.}} & 
\multirow{2}{*}{\shortstack{Q-Sylvan \\ Pauli}} & 
\multirow{2}{*}{\shortstack{Quokka \\Sharp}} & 
\multirow{2}{*}{\shortstack{MQT \\QCEC}} & 
\multirow{2}{*}{\shortstack{Q-Sylvan \\ alter.}} & 
\multirow{2}{*}{\shortstack{Q-Sylvan \\ Pauli}} & 
\multirow{2}{*}{\shortstack{Quokka \\Sharp}} & 
\multirow{2}{*}{\shortstack{MQT \\QCEC}} \\
& & & & & & & & & & & \\
\hline
Portfolio QAOA & 5 & 195 & 236 & 0.18 & 6.66 & 14.20 & \textbf{0.00} & 0.42 & 6.36 & 39.78 & \textbf{0.00} \\
Portfolio QAOA & 6 & 261 & 356 & 1.91 & 37.04 & 123.02 & \textbf{0.01} & 2.02 & 35.05 & > 300 & \textbf{0.01} \\
Portfolio QAOA & 7 & 336 & 481 & 8.95 & 213.99 & > 300 & \textbf{0.02} & 10.52 & 184.41 & > 300 & \textbf{0.03} \\
Portfolio VQE & 5 & 310 & 131 & 0.13 & 5.67 & 20.77 & \textbf{0.00} & 0.16 & 5.77 & 30.19 & \textbf{0.00} \\
Portfolio VQE & 6 & 435 & 151 & 1.20 & 27.86 & > 300 & \textbf{0.01} & 1.27 & 26.97 & > 300 & \textbf{0.01} \\
Portfolio VQE & 7 & 581 & 218 & 1.90 & 150.14 & > 300 & \textbf{0.01} & 0.36 & 133.26 & > 300 & \textbf{0.01} \\
Pricing call & 5 & 240 & 166 & 0.10 & 1.33 & 0.31 & \textbf{0.00} & 0.10 & 1.54 & 0.39 & \textbf{0.00} \\
Pricing call & 7 & 422 & 277 & 0.25 & 21.48 & 63.51 & \textbf{0.01} & 0.28 & 19.34 & 55.49 & \textbf{0.01} \\
Pricing call & 9 & 624 & 396 & 231.28 & > 300 & > 300 & \textbf{0.02} & 228.99 & > 300 & > 300 & \textbf{0.02} \\
Pricing put & 5 & 240 & 192 & 0.12 & 2.20 & 0.37 & \textbf{0.00} & 0.12 & 1.85 & 0.40 & \textbf{0.00} \\
Pricing put & 7 & 432 & 297 & 5.97 & 24.70 & > 300 & \textbf{0.01} & 4.23 & 15.29 & 167.63 & \textbf{0.01} \\
Pricing put & 9 & 654 & 428 & > 300 & > 300 & > 300 & \textbf{0.04} & 252.70 & > 300 & > 300 & \textbf{0.03} \\
QAOA & 7 & 133 & 117 & 0.10 & 1.10 & 0.08 & \textbf{0.01} & 0.25 & 1.02 & 0.31 & \textbf{0.01} \\
QAOA & 9 & 171 & 296 & 1.40 & 2.45 & 0.15 & \textbf{0.02} & 28.79 & 2.38 & 0.12 & \textbf{0.02} \\
QAOA & 11 & 209 & 359 & > 300 & 4.36 & 0.14 & \textbf{0.06} & - & 5.59 & 0.60 & \textbf{0.05} \\
QFT & 2 & 14 & 14 & \textbf{0.00} & 0.01 & 0.00 & 0.00 & 0.01 & 0.02 & 0.00 & \textbf{0.00} \\
QFT & 8 & 176 & 228 & 1.84 & 40.29 & 0.06 & \textbf{0.01} & 8.20 & 43.72 & 0.04 & \textbf{0.01} \\
QFT & 16 & 672 & 814 & > 300 & > 300 & 3.41 & \textbf{0.40} & > 300 & - & 178.31 & \textbf{0.26} \\
QFT entangled & 16 & 690 & 853 & > 300 & > 300 & \textbf{53.32} & 63.23 & > 300 & > 300 & \textbf{0.28} & 25.78 \\
QFT entangled & 32 & 2658 & 3068 & - & - & > 300 & > 300 & - & - & > 300 & > 300 \\
QFT entangled & 64 & 10434 & 9387 & > 300 & > 300 & > 300 & > 300 & > 300 & > 300 & > 300 & > 300 \\
QNN & 2 & 43 & 36 & 0.01 & 0.08 & 0.03 & \textbf{0.00} & 0.01 & 0.08 & 0.03 & \textbf{0.00} \\
QNN & 8 & 319 & 494 & 15.10 & > 300 & > 300 & \textbf{0.03} & 12.83 & > 300 & > 300 & \textbf{0.02} \\
QNN & 16 & 1023 & 2002 & > 300 & > 300 & > 300 & \textbf{28.31} & > 300 & > 300 & > 300 & \textbf{29.07} \\
QPE exact & 16 & 712 & 837 & > 300 & - & > 300 & \textbf{2.36} & > 300 & > 300 & \textbf{0.75} & 3.35 \\
QPE exact & 32 & 2712 & 3276 & > 300 & > 300 & > 300 & > 300 & > 300 & > 300 & > 300 & > 300 \\
QPE exact & 64 & 10552 & 9680 & > 300 & > 300 & > 300 & > 300 & > 300 & > 300 & > 300 & > 300 \\
QPE inexact & 16 & 712 & 848 & - & > 300 & 17.77 & \textbf{4.35} & > 300 & > 300 & 29.12 & \textbf{4.80} \\
QPE inexact & 32 & 2712 & 3179 & > 300 & > 300 & > 300 & > 300 & > 300 & > 300 & > 300 & > 300 \\
QPE inexact & 64 & 10552 & 9695 & > 300 & > 300 & > 300 & > 300 & > 300 & > 300 & > 300 & > 300 \\
Q-walk (no an.) & 6 & 2457 & 3003 & 11.61 & > 300 & > 300 & \textbf{0.03} & 18.43 & > 300 & > 300 & \textbf{0.03} \\
Q-walk (no an.) & 7 & 4761 & 6231 & 202.57 & > 300 & > 300 & \textbf{0.06} & 215.58 & > 300 & > 300 & \textbf{0.07} \\
Q-walk (no an.) & 8 & 9369 & 12486 & > 300 & > 300 & > 300 & \textbf{0.22} & > 300 & > 300 & > 300 & \textbf{0.22} \\
Q-walk (v-chain) & 5 & 417 & 398 & 0.24 & 3.81 & 18.14 & \textbf{0.00} & 0.26 & 3.42 & 17.93 & \textbf{0.00} \\
Q-walk (v-chain) & 7 & 849 & 840 & 1.33 & > 300 & > 300 & \textbf{0.01} & 8.40 & > 300 & > 300 & \textbf{0.01} \\
Q-walk (v-chain) & 9 & 1437 & 1500 & 22.45 & > 300 & > 300 & \textbf{0.04} & > 300 & > 300 & > 300 & \textbf{0.03} \\
Real amp. rand. & 16 & 680 & 679 & > 300 & > 300 & > 300 & \textbf{68.97} & > 300 & > 300 & > 300 & \textbf{58.39} \\
Real amp. rand. & 32 & 2128 & 2215 & > 300 & > 300 & \textbf{57.77} & > 300 & > 300 & > 300 & > 300 & > 300 \\
Real amp. rand. & 64 & 7328 & 7411 & > 300 & > 300 & > 300 & > 300 & > 300 & > 300 & > 300 & > 300 \\
Routing & 2 & 43 & 29 & 0.02 & 0.08 & 0.02 & \textbf{0.00} & 0.02 & 0.08 & 0.02 & \textbf{0.00} \\
Routing & 6 & 135 & 142 & 0.05 & 6.59 & 0.06 & \textbf{0.01} & 0.06 & 6.63 & 0.31 & \textbf{0.01} \\
Routing & 12 & 273 & 409 & - & > 300 & 40.01 & \textbf{0.29} & - & > 300 & \textbf{0.16} & 0.28 \\
SU2 rand. & 16 & 744 & 378 & - & > 300 & \textbf{2.83} & 24.20 &   &   &   &   \\
SU2 rand. & 32 & 2256 & 762 & > 300 & > 300 & > 300 & > 300 &   &   &   &   \\
SU2 rand. & 64 & 7584 & 1530 & > 300 & > 300 & > 300 & > 300 &   &   &   &   \\
TSP & 4 & 225 & 86 & 0.08 & 1.56 & 2.63 & \textbf{0.00} & 0.15 & 1.39 & 2.94 & \textbf{0.00} \\
TSP & 9 & 550 & 315 & 238.04 & > 300 & > 300 & \textbf{0.03} & 254.58 & > 300 & > 300 & \textbf{0.07} \\
TSP & 16 & 1005 & 623 & > 300 & > 300 & > 300 & \textbf{7.78} & > 300 & > 300 & > 300 & \textbf{8.13} \\
Two-local rand. & 16 & 680 & 679 & > 300 & > 300 & > 300 & \textbf{75.84} & > 300 & > 300 & > 300 & \textbf{59.00} \\
Two-local rand. & 32 & 2128 & 2215 & > 300 & > 300 & \textbf{57.62} & > 300 & > 300 & > 300 & > 300 & > 300 \\
Two-local rand. & 64 & 7328 & 7411 & > 300 & > 300 & > 300 & > 300 & > 300 & > 300 & > 300 & > 300 \\
VQE & 5 & 83 & 83 & $\times$ & $\times$ & 0.02 & \textbf{0.00} & 0.06 & 0.70 & 0.03 & \textbf{0.00} \\
VQE & 10 & 168 & 221 & 150.22 & > 300 & 15.96 & \textbf{0.07} & > 300 & > 300 & \textbf{0.05} & 0.06 \\
VQE & 15 & 253 & 349 & > 300 & - & \textbf{0.30} & 37.44 & > 300 & - & 133.53 & \textbf{61.72} \\
W-state & 16 & 271 & 242 & > 300 & 4.51 & 0.79 & \textbf{0.01} & > 300 & 4.91 & 0.31 & \textbf{0.01} \\
W-state & 32 & 559 & 498 & > 300 & 33.79 & 4.18 & \textbf{0.04} & > 300 & > 300 & 1.81 & \textbf{0.01} \\
W-state & 64 & 1135 & 1010 & > 300 & 221.67 & \textbf{24.65} & > 300 & > 300 & - & 11.59 & \textbf{1.71} \\
\end{tabular}

    }
\end{table}
}{}

\end{document}